# Enhanced Inter-Cell Interference Coordination Challenges in Heterogeneous Networks

David López-Pérez[1], İsmail Güvenç[2], Guillaume de la Roche[3], Marios Kountouris[4], Tony Q.S. Quek[5], Jie Zhang[6]
[1] King's College London, London, UK, Email: david.lopez@kcl.ac.uk
[2] DOCOMO Communications Laboratories USA, Inc., Palo Alto, CA, Email: iguvenc@docomolabs-usa.com
[3] University of Bedfordshire, Luton, UK, Email: guillaume.delaroche@beds.ac.uk
[4] SUPELEC, Gif-sur-Yvette, France, Email: marios.kountouris@supelec.fr
[5] Institute for Infocomm Research, A*STAR, Connexis, Singapore, Email: qsquek@i2r.a-star.edu.sg
[6] University of Sheffield, Sheffield, UK, Email: jie.zhang@sheffield.ac.uk

*Abstract*—3GPP LTE-Advanced has started a new study item to investigate Heterogeneous Network (HetNet) deployments as a cost effective way to deal with the unrelenting traffic demand. HetNets consist of a mix of macrocells, remote radio heads, and low-power nodes such as picocells, femtocells, and relays. Leveraging network topology, increasing the proximity between the access network and the end-users, has the potential to provide the next significant performance leap in wireless networks, improving spatial spectrum reuse and enhancing indoor coverage. Nevertheless, deployment of a large number of small cells overlaying the macrocells is not without new technical challenges. In this article, we present the concept of heterogeneous networks and also describe the major technical challenges associated with such network architecture. We focus in particular on the standardization activities within the 3GPP related to enhanced inter-cell interference coordination.

## I. Introduction

With more than one billion wireless subscribers today and predictions for this number being tripled over the next five years, the wireless industry is confronted with an increasing demand for ubiquitous wireless coverage and larger data rates. The exponential traffic growth in broadband wireless networks is a well established fact and this unprecedented trend is accelerated by the proliferation of advanced user terminals and bandwidth-greedy applications, e.g., mobile TV, file transfer. In order to support this galloping demand for data traffic, 3rd Generation Partnership Project (3GPP) Long Term Evolution (LTE) Release 8 is under field trial by most cellular operators. This standard offers significant advantages with respect to its predecessor High Speed Packet Access (HSPA), for instance, higher spectral efficiency, lower latency due to its flat all-Internet Protocol (IP) architecture, and larger throughputs [1]. However, the performance of Release 8 does not meet the International Mobile Telecommunications (IMT)-Advanced requirements for the fourth generation of mobile networks defined by the International Telecommunication Union (ITU). Thus, in order to meet such requirements (e.g., downlink data rates of up to 100 Mbps and 1 Gbps for mobile and nomadic users, respectively), LTE-Advanced, i.e., LTE Release 10, is currently under standardization.

In order to enhance the performance of the overall network, LTE-Advanced proposes the use of advanced technologies [2]. For instance, Carrier Aggregation (CA) allows the concurrent utilization of different frequency carriers, hence efficiently increasing the bandwidth that can be allocated to end-users. Another trend is the enhancement of multi-antenna techniques, where using Multiple-Input Multiple-Output (MIMO) systems with up to $8 \times 8$ antenna arrays has gained significant attention. Coordinated Multi-Point (CoMP) transmission and reception, where multiple cells are able to coordinate their scheduling or transmission to serve users with adverse channel conditions, is also envisioned to notably mitigate outages at the cell-edge. However, all these advanced technologies do not allow significant enhancements as they are reaching theoretical limits. Such techniques may not always work well either, especially under low Signal-to-Interference Plus Noise Ratio (SINR) conditions, where received powers are low due to attenuation, e.g., residential/office scenarios.

In order to overcome these issues and provide a significant network performance leap, heterogeneous networks (HetNets) have been introduced in the LTE-Advanced standarization. A HetNet uses a mix of macrocells, Remote Radio Heads (RRHs) and low power nodes such as picocells, femtocells, and relays in order to bring the network closer to end-users. In that way, radio link quality can be enhanced due to the reduced distance between transmitter and receiver, and the larger number of cells allows for more efficient spectrum reuse and therefore larger data rates. As a result, HetNets are expected to be one of the major performance enhancement enablers of LTE-Advanced.

## II. HetNets

A HetNet is a network consisting of infrastructure points with various wireless access technologies, each of them having different capabilities, constraints, and operating functionalities. Specifically, in LTE-Advanced, multi-tier network roll-outs, involving RRHs, picocells, femtocells, as well as relay stations underlaying the existing macrocellular layout are envisaged. These low-power overlaid Base Stations (BSs) can be either operator deployed or user deployed, and may coexist in the same geographical area, potentially sharing the same spectrum. Deploying such small cells aims at offloading the macrocells, improving indoor coverage and cell-edge user performance, and boosting spectral efficiency per area unit via spatial reuse. They can be deployed with relatively low network overhead, and have high potential for reducing the energy consumption of future wireless networks. Also, this new palette of low-power 'miniature' BSs requires little or no upfront planning and lease costs, therefore drastically reducing the Operational (OPEX) and Capital (CAPEX) Expenditures of networks [3].

According to Table I, we provide details of the different elements of HetNets as follows:

- Macrocellular networks consist of conventional operator-installed BSs, providing open public access and a wide area coverage typically on the order of few kilometers. In LTE, they are also called enhanced NodeBs (eNBs). Usually destined to provide guaranteed minimum data rate under maximum tolerable delay and outage constraints, macrocells typically emit up to $46\,\text{dBm}$, serving thousands of customers and using a dedicated backhaul.
- Picocells are low-power, operator-installed cell towers with the same backhaul and access features as macrocells. They are usually deployed in a centralized way, serving few tens of users within a radio range of $300\,\text{m}$ or less, and have a typical transmit power range from $23\,\text{dBm}$ to $30\,\text{dBm}$. Picocells are mainly utilized for capacity and outdoor or indoor coverage infill, i.e., in environments with insufficient macro penetration (e.g., office buildings).
- Femtocells, also known as home BSs or home eNBs, are low-cost, low-power, user-deployed access points, offloading data traffic using consumers' broadband connection (Digital Subscriber Line (DSL), cable, or fiber), and serving a dozen of active users in homes or enterprises. Typically, the femtocell range is less than $50\,\text{m}$ and its transmit power less than $23\,\text{dBm}$. They operate in open or restricted (Closed Subscriber Group (CSG)) access.
- Relays are usually operator-deployed access points that route data from the macro BS to end-users and vice versa. They are positioned so as to increase signal strength and to improve reception in poor coverage areas and dead-spots in the existing networks (e.g., cell edges, tunnels). They can operate in transparent or non-transparent modes (e.g., IEEE 802.16m), with little to no incremental backhaul expense and with similar transmit power as picocells.
- RRH are compact-size, high-power and low-weight units, which are mounted outside the conventional macro BS, and connected to it generally through a fiber optic cable, thus creating a distributed BS. The central macro BS is in charge of control and baseband signal processing, Moving some radio circuitry into the remote antenna, RRHs eliminate power losses in the antenna cable and reduce power consumption. They also enhance flexibility to network deployments for operators that face site acquisition challenges and/or physical limitations them.

HetNets entail a significant paradigm shift transitioning from 'traditional', centralized macrocell/microcell approaches to more autonomous, uncoordinated, and intelligent roll-outs.

TABLE I
SPECIFICATION OF DIFFERENT ELEMENTS IN HETNET

| Types of nodes | Transmit Power | Coverage | Backhaul |
|---|---|---|---|
| Macrocell | 46 dBm | Few km | S1 interface |
| Picocell | 23-30 dBm | <300 m | X2 interface |
| Femtocell | <23 dBm | <50 m | Internet IP |
| Relay | 30 dBm | 300 m | Wireless |
| RRH | 46 dBm | Few km | Fiber |

However, this paradigm shift, which can be seen as an excellent opportunity for enhancements, also introduces challenges. For instance, although the advantage of deploying femtocells is supported by recent studies suggesting that $50\,\%$ of all voice calls and more than $70\,\%$ of data traffic originate indoors, cross-tier interference and traffic load variability may become a barrier to a successful deployment of this type of network. Hence, HetNets bring into play significant technical issues and raise substantial challenges that are presented in the next section.

## III. TECHNICAL CHALLENGES

Self-organization, backhauling, handover, and interference are identified here as the key technical challenges facing HetNets.

### A. Self-organization

Since some cells such as picocells and femtocells will be user-deployed without operator supervision, their proper operation highly depends on their self-organizing features [4]. The self-organizing capability of HetNets can be generally classified into three processes:

- *Self-configuration*, where newly deployed cells are automatically configured by downloaded software before entering into the operational state.
- *Self-healing*, where cells can automatically perform failure recovery or execute compensation mechanisms whenever failures occur.
- *Self-optimization*, where cells constantly monitor the network status and optimize their settings to improve coverage and reduce interference.

The deployment of self-organizing HetNets is an intricate task due to the various type of coexisting cells and the increasing number of network parameters that need to be considered. The random, uneven and time-varying nature of user arrivals and their resulting traffic load also exacerbate the difficulties associated with deploying a completely self-organized HetNet.

### B. Backhauling

Backhaul network design will be a major issue because of the complex topology of the various type of coexisting cells. For instance, the deployment of picocells will require access to utility infrastructure with power supply and wired network backhauling, which may be potentially expensive. Femtocells, which in contrast have relatively lower backhauling costs, may face difficulties in maintaining Quality of Service (QoS) since backhauls rely on consumers' broadband connections. Hence, operators need to plan HetNet backhaul carefully to identify the most cost effective and QoS guaranteed solution. Such a solution is likely to be a mixture of both wireless and wired backhaul technologies, in which some cells may have dedicated interfaces to the core network, some other cells may form a cluster to aggregate and forward the traffic to the core, and other cells may rely on relays as an alternative interface.

### C. Handover

Handovers are essential in order to provide a seamless uniform service when users move in or out of the cell coverage. Furthermore, handovers are efficient for traffic load balancing, by shifting users at the border of adjacent/overlapping cells from the more congested cells to the less congested ones. Nevertheless, this comes at the expense of system overhead, which is likely to be significant in HetNets due to the large number of small cells and the different types of backhaul links available for each type of cell. In addition, the probability of handover failure increases the probability of user outage [4].

### D. Interference

Unlike traditional single-tier cellular networks, in HetNets, the cross-tier and intra-tier interference problems are significantly challenging due to the following reasons:

1) The backhaul network supporting different types of cells may have different bandwidth and delay constraints. For instance, femtocells are unlikely to be connected directly to the core network and thus only limited backhaul signaling for interference coordination is possible.
2) The restricted access control associated with picocells and femtocells may lead to strong interference scenarios in both uplink and downlink since users may not handover to the nearest cells.
3) The self-organizing capability of cells also requires continuous sensing and monitoring of the radio environment around them in order to dynamically and adaptively mitigate or avoid interference.

With interference remaining undebatedly the major challenge, in the sequel we focus on this topic.

## IV. INTERFERENCE RELATED ISSUES IN HETNETS

Within operator-deployed cells like macrocells and relays, interference may be mitigated via frequency reuse schemes e.g., important is the planning of both relay and direct links. However, since these reuse schemes may reduce spatial reuse, subchannels used in a cell are banned in neighboring ones, the tendency is to drop their use and target to universal reuse, where all cells have the potential to use all available resources, taking varying traffic load and channel conditions into account.

Also, roll-outs of user-deployed cells overlying macrocells will create new cell boundaries, in which end-users will suffer from strong inter-cell interference, degrading the performance of the overall cellular network. In this section, the principal interference scenarios in HetNets will be reviewed, followed by a discussion on Inter-Cell Interference Coordination (ICIC) techniques specified in LTE standardization to address them.

Note that due to space limitations, and the fact that femtocells pose a significant challenge to the proper operation of a HetNet due to their unplanned deployment and inter-cell interference characteristics, the rest of this paper is focused on the macrocell interactions with femtocells and picocells.

### A. Sources of Interference

In addition to the large number of created cell boundaries, the interference problem in HetNets is especially challenging due to following reasons:

*1) Unplanned deployment:* Low-power nodes such as femtocells are typically deployed in an ad hoc manner by users. They can even be moved or switched on/off at any time. Hence, traditional network planning and optimization becomes inefficient because operators do not control neither the number nor the location of these cells. This motivates the need for new decentralized interference avoidance schemes that operate independently in each cell, utilizing only local information, whereas achieving an efficient solution for the entire network.

*2) CSG access:* The fact that some cells may operate in CSG mode, in which cell access is restricted and nonsubscribers are thus not always connected to the nearest BS, originates significant cross-tier interference components [5]. Fig. 1 depicts a challenging scenario for ICIC, in which different nonsubscribers walk nearby houses hosting a CSG femtocell. In the Uplink (UL), nonsubscriber (a) transmit at high power to compensate for the path losses to its far serving macrocell, jamming the UL of the nearby CSG femtocell(s). In the Downlink (DL), a CSG femtocell interferes the DL reception of non-subscriber (b) connected to the far macrocell. Hence, this DL Macrocell User (MUE) becomes a victim user.

*3) Power difference between nodes:* Picocells and relays usually operate in open access mode, meaning that all users of a given operator can access to them. Open access helps to minimize DL interference as end-users always connect to the strongest cell, thus avoiding the CSG interference issue. However, in HetNets, being attached to the cell that provides the strongest DL Received Signal Strength (RSS)may not be the best strategy since users will tend to connect to macrocells, and not to those cells being at the shortest path loss distance. This is due to the large difference in transmission power between macrocells and low-power nodes. In that way, traffic load will be unevenly distributed, thus overloading macrocells.

Moreover, due to this server selection procedure in the DL, users connected to macrocells will severely interfere all low-power nodes located in their vicinity in the UL. Fig. 1 (c) illustrates how a user connected to a macrocell, which provides the best DL RSS, jams a nearby picocell UL. Note that due to lower path loss, this MUE would transmit with a much less UL power if it was associated with the picocell. This will allow load balancing and UL interference mitigation, improving network performance.

*4) Range expanded users:* To address the problems arising due to the power difference between the nodes in HetNets, new cell selection methods that allow user association with cells that provide a weaker DL pilot signal quality are necessary. An approach under investigation is that of range expansion [6], in which an offset is added to the picocell's (or relay's) RSS in order to increase its DL coverage footprint (see Fig. 1 (d)). Even though range expansion significantly mitigates cross-tier interference in the UL, this comes at the expense of reducing the DL signal quality of those users in the expanded region. Such users may suffer from DL SINRs below $0\,\text{dB}$ since they are connected to cells that do not provide the best DL RSS (see Fig. 1(d)).

### B. Inter-Cell Interference Coordination

The interference problems summarized above may significantly degrade the overall HetNet performance, which requires the use of ICIC schemes to guarantee its proper operation. In such schemes, special attention should be given to the mitigation of inter-cell interference in the control channels. User Equipments (UEs) may declare radio link failure under severe interference, and experience service outage due to the unreliable DL control channels.

Moreover, it is essential that UEs are able to sense, detect, and report information to their servers concerning potential interfering cells being present in their vicinity. Then, the UE serving cell in collaboration with the potential interferers will coordinate their resource allocation in terms of power, frequency, and time to enhance network capacity and mitigate user outages.

In order to facilitate this coordination between HetNet cells, information messages need to be exchanged among them. Macrocells are connected to picocells and relays through the X2 interface. The ICIC messages defined in Release 8 that can be exchanged via the X2 interface can be listed as follows [7]:

- *Relative Narrowband Transmit Power (RNTP) Indicator:* For DL transmissions, RNTP indicator transmitted by a certain cell is used to inform the neighboring cells on whether the transmit power for specified Resource Blocks (RBs) will be set below a certain threshold value.
- *Overload Indicator (OI):* For uplink transmissions, average interference plus thermal noise power measurements for each RB is exchanged between different cells.
- *High Interference Indicator (HII):* A certain cell informs neighboring cells that uplink transmission of one of its cell-edge users will be scheduled in the near future, and neighboring cells may abstain from scheduling their own cell-edge users or high powers in the specified RBs.

While such messages exchanged over the X2 interface can help alleviating the dominant interference scenarios for macrocells, picocells, and relays, they are not available for femtocells, hence necessitating new interference coordination approaches. One possible solution enabling macro/femto coordination is the exchange of such information messages via the backhaul. However, because the wireline backhaul of femtocells may not be owned by the network operator, delay issues may appear. To overcome this issue, the exchange of messages between macros and femtos through the wireless broadcast channels or the use of UEs for relaying data between neighboring cells are being investigated [8]. For instance, victim MUEs can be determined by the macrocell by utilizing the measurement reports of the MUEs, and their identity may be signaled by the macrocell to the femtocell through the backhaul connection. Similarly, another approach in which low-power nodes are able to detect macrocell users and take actions to prevent outages is under investigation [9]. This option will minimize the exchange of data between cells at the expense of having more sophisticated and expensive hardware at the small cells.

## V. STANDARDIZATION FOR HETNET eICIC

The ICIC methods specified in Release 8 and Release 9 do not specifically consider HetNet settings and may not be effective for dominant HetNet interference scenarios (Fig. 1). In order to address such dominant interference scenarios, enhanced Inter-Cell Interference Coordination (eICIC) techniques are recently being developed for Release 10, which can be grouped under three major categories according to [9]:

- Time-domain techniques.
- Frequency-domain techniques.
- Power control techniques.

These approaches will be reviewed in more detail below. Unless otherwise stated, femtocell eICIC is considered because of its priority in 3GPP.

### A. Time-Domain Techniques

In time-domain eICIC methods, transmissions of the victim users are scheduled in time-domain resources (e.g., subframe, or OFDM symbol), where the interference from other nodes is mitigated. They can be classified into two categories as follows [10]:

*1) Subframe Alignment:* When the subframes of macro eNB and home eNB are aligned as in Fig. 2(a), their control and data channels overlap with each other. Therefore, in order not to interfere to the control channel of MUEs, control channel eICIC needs to be implemented at the femtocells. One possible way to achieve this is to use so-called *Almost Blank Subframes* (ABSFs) at femtocells, as shown in Fig. 2(c). In the ABSFs, no control or data signals, but only reference signals are transmitted[1]. When there are MUEs in the vicinity of a femtocell, they can be scheduled within the subframes overlapping with the ABSFs[2] of the femtocell, which significantly mitigates cross-tier interference.

Similar eICIC approach using ABSFs can also be used to mitigate interference problems in picocells (and relays) that implement range-expansion. When no interference coordination is used for range-expanded picocell users (Fig. 2(b)), they observe large DL interference from the macrocell. The interference problem can be mitigated through using ABSFs at the macrocell, and scheduling range-expanded picocell users within the subframes that are overlapping with the ABSFs of the macrocell.

---

[1] Note that femtocells still need to transmit reference signals in the coordinated subframes, which occupy a limited portion of the whole subframe. These reference signals may still cause some severe interference problems in dominant interference settings [10].

[2] Note that while only the even subframes are configured as ABSFs in this paper, different ABSF patterns are also possible. For example, current patterns considered in 3GPP have ABS duty cycles of 1/8, 2/8, 3/8, and 3/20 [11].

4*2) OFDM Symbol Shift:* In the second category of time-domain eICIC methods, subframe boundary of home eNB is shifted by a number of OFDM symbols with respect to the subframe boundary of macro eNB in order to prevent the overlap between the control channels of femtocell and macrocell signals [10]. However, there still exists interference from the data channels of femtocell users to the control channels of macrocell users. Two possible solutions to address this problem are 1) shared-channel symbol muting, and 2) consecutive subframe blanking at femtocells. In the first approach, the OFDM symbols that overlap with the control channel of the victim MUEs are muted. In the second approach, the subframes of femtocell that overlap with the control channel of MUEs are configured as ABSF.

### B. Frequency-Domain Techniques

In frequency-domain eICIC solutions, control channels and physical signals (i.e., synchronization signals and reference signals) of different cells are scheduled in reduced bandwidths in order to have totally orthogonal transmission of these signals at different cells. While frequency-domain orthogonalization may be achieved in a static manner, it may also be implemented dynamically through victim UE detection.

For instance, victim MUEs can be determined by the macro eNBs by utilizing the measurement reports of the MUEs, and their identity may be signaled by the macro eNB to the home eNB(s) through the backhaul. Alternatively, victim MUEs may also be sensed by the home eNBs.

### C. Power-Control Techniques

One last approach that has been heavily discussed in 3GPP for handling dominant interference scenarios is to apply different power control techniques at femtocells. While reducing the radiated power at a femtocell also reduces the total throughput of femtocell users, it may significantly improve the performance of victim MUEs.

Let $P_{\max}$ and $P_{\min}$ denote the maximum and minimum home eNB transmit powers, respectively, $P_M$ denote the received power from the strongest co-channel macro eNB at a home eNB, $\alpha$ and $\beta$ denote two scalar power control variables. Then, different DL power control approaches at femtocells can be listed as follows (all values are in dBm) [9]:

1) *Strongest macro eNB received power at a home eNB:* The femtocell transmission power can be written as $P_{\text{tx}} = \max\big(\min(\alpha P_M + \beta, P_{\max}), P_{\min}\big)$.
2) *Path loss between a home eNB and MUE:* The home eNB transmission power can be set as $P_{\text{tx}} = \text{med}\big(P_M + P_{\text{ofst}}, P_{\max}, P_{\min}\big)$, where the power offset is defined by $P_{\text{ofst}} = \text{med}\big(P_{\text{ipl}}, P_{\text{ofst}-\max}, P_{\text{ofst}-\min}\big)$, with $P_{\text{ipl}}$ denoting a power offset value that captures the indoor path loss and the penetration loss between home eNB and the nearest MUE, and $P_{\text{ofst}-\max}$ and $P_{\text{ofst}-\min}$ denote the minimum and maximum values of $P_{\text{ofst}}$, respectively.
3) *Objective SINR of HUE:* In this approach, the received SINRs of home eNB users (HUEs) are restricted to a target value and transmit power at a femtocell is reduced appropriately to achieve this target SINR using the following expression: $P_{\text{tx}} = \max\big(P_{\min}, \min(\widehat{PL} + P_{\text{rec,HUE}}, P_{\max})\big)$, where $P_{\text{rec,HUE}} = 10\log_{10}\big(10^{I/10} + 10^{N_0/10}\big) + \text{SINR}_{\text{tar}}$, with $I$ being the interference detected by the served UE, $N_0$ is the background noise power, $\text{SINR}_{\text{tar}}$ is the target SINR for the HUE, and $\widehat{PL}$ is the path loss estimate between the home eNB and the HUE.
4) *Objective SINR of MUE:* The goal of this approach is to guarantee a minimum SINR at the MUEs, and the home eNB transmit power is given by $P_{\text{tx}} = \max\big(\min(\alpha P_{\text{SINR}} + \beta, P_{\max}), P_{\min}\big)$, where $P_{\text{SINR}}$ is the SINR of the MUE considering only the nearest femtocell interference.

## VI. eICIC Performance Analysis

In this section, the DL of an LTE-Advanced HetNet is simulated to test the different eICIC schemes presented above. The scenario (Fig. 3) under scrutiny is a residential area of size $300\,\text{m} \times 300\,\text{m}$ in Luton (UK), containing 400 dwelling houses of which 63 were selected to host a CSG femtocell (if we assume 3 network operators with equal customer share, this corresponds to an approximate 50 % femto penetration). The scenario is also covered by one macrocell located 200 m south and 200 m east from the scenario's center (i.e., outside of Fig. 3), and one picocell deployed at the macrocell edge. Both macrocell and picocell operate in open access.

Eight mobile users, utilizing a Voice over IP (VoIP) service, move along predefined paths according to a pedestrian model of mean speed $1.1\,\text{m/s}$. Meanwhile, the picocell and the femtocells are fully loaded and therefore utilize all subcarriers. The cell power is uniformly distributed between subcarriers, and a pedestrian user carrying a VoIP service is considered to fall in outage if it cannot receive control data (i.e., user SINR is smaller than $-4\,\text{dB}$ for a time interval of $200\,\text{ms}$).

### A. Macrocell - Femtocell Interaction

Fig. 4 illustrates the SINR of a pedestrian user when passing by the front door of two different houses hosting a femtocell. It can be seen that when no action is taken at the femtocells (no eICIC), the SINR of the pedestrian user significantly falls due to the cross-tier interference, thus resulting in user outage. On the other hand, when eICIC is applied, the MUE SINR recovers and outages vanish. In this case, an eICIC action is triggered by the macrocell in the femtocells when MUEs report low signal quality using channel quality indicators, i.e., user SINR smaller than -3 dB.

The ABSF eICIC time method provides the best MUE protection since those subframes overlapping with the ABSFs of femtocells are not interfered. The different eICIC power methods presented in Sec. V-C result on the other hand in distinct levels of signal quality protection for the victim MUE. The behavior of these eICIC techniques depends on their nature and tuning, but there is always a tradeoff between the performance of both victim MUE and aggressing femtocell. Table II, where the simulation results are given, shows this. The larger the average sum throughput of the eight MUEs,

TABLE II
PERFORMANCE COMPARISON (600 SEC SIMULATION)

| eICIC methods | Number of macro-pico HOs | Number of PUE outages | Number of MUE outages | Average eICIC TP gain at a femto [Mbps] | Average sum TP of pedestrian users [kbps] | Average sum TP of femtocell tier [Mbps] | $\frac{eICIC\ actions}{femto \cdot 10\ min}$ |
|---|---|---|---|---|---|---|---|
| no eICIC | 5 | 5 | 267 | 73.32(100 %) | 156.03 | 3974.25 | - |
| eICIC time | 5 | 0 | 0 | 0 (0 %) | 2158.82 | 2990.50 | 14.81 |
| eICIC power 1 $\alpha = 1, \beta = 60$dB | 5 | 0 | 0 | 11.02 (15.03 %) | 1937.26 | 3153.88 | 14.81 |
| eICIC power 1* $\alpha = 1, \beta = 75$dB | 5 | 0 | 25 | 46.49 (63.41 %) | 1139.20 | 3725.88 | 56.23 |
| eICIC power 2 | 5 | 0 | 0 | 34.49 (47.03 %) | 1499.30 | 3558.75 | 20.80 |
| eICIC power 3 $\text{SINR}_{\text{tar}}^{\text{FUE}} = 0$dB | 5 | 0 | 0 | 22.55 (30.75 %) | 1626.61 | 3333.75 | 17.47 |
| eICIC power 3* $\text{SINR}_{\text{tar}}^{\text{FUE}} = 5$dB | 5 | 0 | 19 | 33.74 (46.02 %) | 1281.21 | 3520.75 | 47.52 |
| eICIC power 4 $\text{SINR}_{\text{tar}}^{\text{MUE}} = 5$dB | 5 | 0 | 0 | 33.74 (66.05 %) | 1183.35 | 3751.13 | 39.78 |

* This eICIC method has not been properly tuned to avoid the number of user outages. They are given for comparison purposes.

the smaller the average sum throughput of the femtocell tier. In addition, in the fifth column: 'eICIC TP gain at a femto', the average throughput of a femtocell is given when an eICIC is triggered on it. The ABSF eICIC time scheme provides the best MUE performance at the expense of 'switching off' the femtocell (no data is carried in ABSFs). On the contrary, taking no action at the femtocells results in the worst MUE protection but in the best femtocell throughput performance.

Furthermore, in Table II, it can also be observed how the performance of power methods (1) and (3) highly depends on the tuning of their parameters, i.e., $\alpha$, $\beta$, and $\text{SINR}_{\text{tar}}$. If they are not fine tuned a large number of outages occurs. On the contrary, power methods (2) and (4) do not depend on this fine tuning, since they are able to adapt to each victim MUE situation considering either its path loss or its SINR. Because of this, power methods (2) and (4) can offer a 'tailored' protection for the victim MUE, avoiding outages while recovering the maximum throughput at each femtocell.

The last column of Table II indicates the average number of eICIC actions triggered in each femtocell every 10 minutes. When the eICIC power methods are not fine tuned, more than one eICIC action are triggered per femtocell to avoid an MUE call drop. This is because the eICIC actions cannot recover the MUE SINR to an adequate value. Hence, when a new channel quality indicator is fed back from the MUE to its macrocell, new eICIC actions are executed. When this case takes place, the MUE normally incurs outage. In addition, it must be noted that power methods (2) and (4) also cast more eICIC actions. This is because the path losses and SINRs of victim MUEs continuously change when they move closer to the femtocells, and thus femtocells need to update their transmission power in order to prevent MUE outage. Therefore, it can be stated that this tailored protection comes at the expense of larger backhaul signaling between cells.

*B. Macrocell - Picocell Interaction*

When pedestrian users are connected to the picocell in its range-expanded region, the macrocell uses ABSFs in order to mitigate macrocell to picocell user DL inter-cell interference. A 10 dB offset was added to the picocell's RSS in the DL to further increase its footprint and create the expanded region. Table II indicates that macro/pico handovers accorded 5 times. Results also show that due to the utilization of ABSFs at the macrocell, picocell user outages were avoided. When no eICIC is carried out at the macrocell, handovers resulted in outage.

VII. CONCLUSION

HetNets have the potential to significantly boost network performance, benefiting from transmitter-to-receiver distance reduction and enabling better spatial reuse of the spectrum. This article has identified the major advantages of HetNets, as well as their technical challenges and research problems. Particular attention has been given to the avoidance of cross-tier interference due to its crucial role in proper operation of multi-tier networks. Furthermore, the main eICIC techniques currently under discussion in 3GPP have been evaluated through realistic system-level simulations.

**David López-Pérez** received his Bachelor and Master degrees in Telecommunication from Miguel Hernandez University, Elche (Spain) in 2003 and 2006, respectively. He joined VODAFONE SPAIN in 2005, working at the Radio Frequency Department in the area of network planning and optimization. He took up a research scholarship at the Cork Institute of Technology in Ireland in 2006, where he was researching indoor positioning systems and sensor networks. In 2007, he joined the Centre for Wireless Network Design (CWiND) at the University of Bedfordshire as PhD Marie-Curie Fellow, where he was involved in different FP6 and FP7 European projects, as well as EPSRC. Since August 2010, he is a postdoctoral research associate at King's College London at the Centre for Telecommunication Research (CTR) working on heteregeneous networks.

**İsmail Güvenç** received his Ph.D. degree in Electrical Engineering from University of South Florida, Tampa, FL, in 2006. He was with Mitsubishi Electric Research Labs in Cambridge, MA, in 2005, and since 2006, he has been with DOCOMO USA Labs, Palo Alto, CA, working as a research engineer. Dr. Guvenc is an associate editor for IEEE Communications Letters since 2010, and he has recently served as a guest editor for EURASIP Journal on Advances in Signal Processing and IEEE Journal on Selected Topics in Signal Processing. He co-authored/co-edited two books on short-range wireless communications and positioning. Dr. Guvenc holds 4 U.S. patents, and has another 20 pending U.S. patent applications.

**Guillaume de la Roche** is a senior research fellow at the Centre for Wireless Network Design (CWiND), United Kingdom, since 2007. Before he was with Infineon (2001-2002, Germany), Sygmum (2003-2004, France) and CITI Laboratory (2004-2007, France). He was also a visiting researcher at DOCOMO-Labs (2010, USA) and AxisTeknologies (2011, USA). He holds a Dipl-Ing from CPE Lyon, and a M.Sc and Ph.D. from INSA Lyon. He is the PI of European FP7 project CWNetPlan.

**Marios Kountouris** (S'04-M'08) received the Dipl.-Ing. in ECE from the National Technical University of Athens, Greece, in 2002 and the M.Sc. and Ph.D. degrees in Electrical Engineering from the Ecole Nationale Supérieure des Télécommunications (Telecom ParisTech), France, in 2004 and 2008, respectively. His doctoral research was carried out at Eurecom Institute, France, funded by FT - Orange Labs, France. From February 2008 to May 2009, he has been with the Department of ECE at the University of Texas at Austin, USA, as a postdoctoral research associate, working on wireless ad hoc networks under DARPA's ITMANET program. Since June 2009 he has been with the Department of Telecommunications at SUPELEC (Ecole Suprieure d'Electricité), France where he is currently an assistant professor. He also received the Best Paper Award in IEEE Globecom 2009. He is a member of IEEE and a professional Engineer of the Technical Chamber of Greece.

PLACE PHOTO HERE

**Tony Q.S. Quek** received the B.E. and M.E. degrees from Tokyo Institute of Technology. At Massachusetts Institute of Technology, he earned the Ph.D. in Electrical Engineering and Computer Science. Since 2008, he has been with the Institute for Infocomm Research, where he is currently a Principal Investigator. He received the Singapore Government Scholarship in 1993, Tokyu Foundation Fellowship in 1998, the A*STAR National Science Scholarship in 2002, and the JSPS Invited Fellow for Research in Japan, 2011. He was honored with the 2008 Philip Yeo Prize for Outstanding Achievement in Research, and the IEEE Globecom 2010 Best Paper Award.

PLACE PHOTO HERE

**Jie Zhang** is a Professor at the Communications Group, the Department of Electronic and Electrical Engineering, University of Sheffield (www.sheffield.ac.uk). Before taking the Chair in Wireless Systems at Sheffield in Jan. 2011, from 1997 to 2010, he had been with University of Bedfordshire, Oxford University, Imperial College London and University College London etc. His research interests are focused on radio network planning and optimisation. Since 2003, as the Principal Investigator, he has been awarded 17 projects worth over 4.0 million pounds (his share) by the EPSRC, the European Commission (FP6/FP7) and the industry etc. He was/is a Co-Investigator of two EPSRC-funded projects on femtocell (B)4G mobile communications. Since 2006, he has published over 100 papers in referred journals and conferences, over 10 of which have been widely cited. He is an author of the book - "Femtocells: Technologies and Deployment" (Wiley, Jan. 2010).

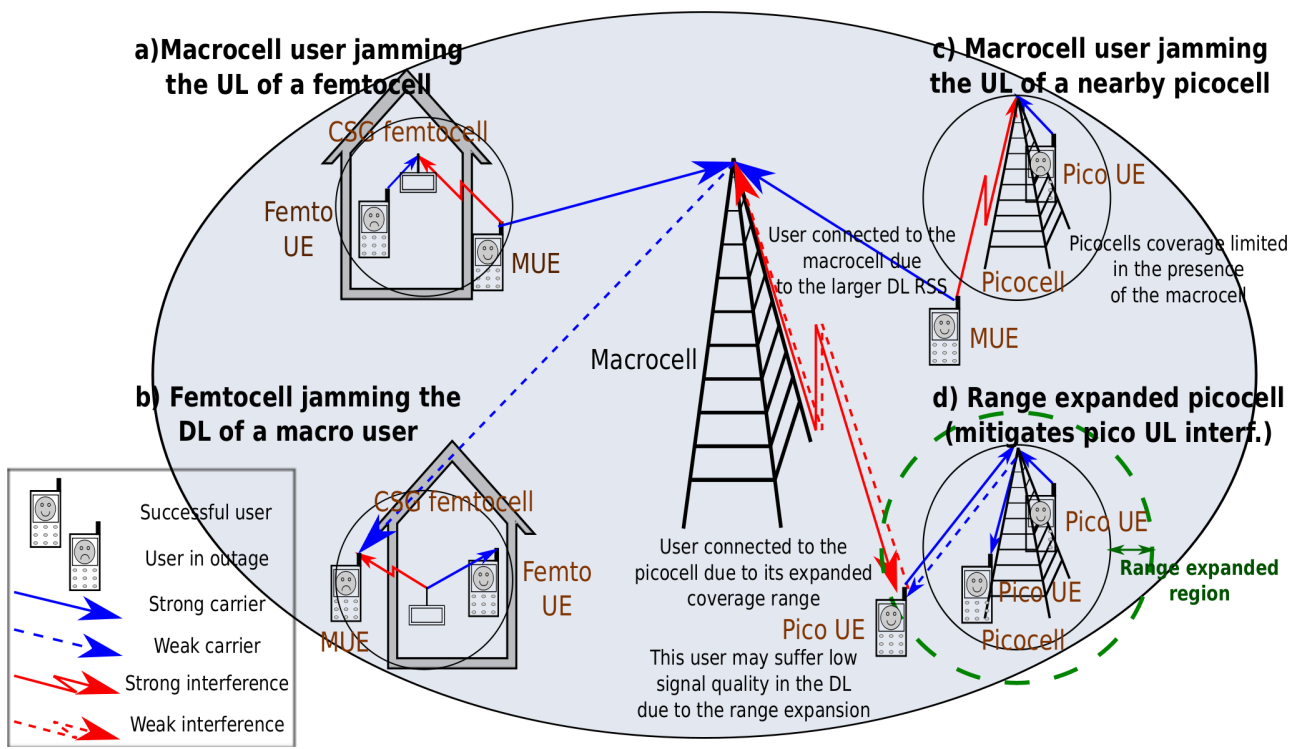

Fig. 1. Dominant DL and UL cross-tier interference scenarios in HetNets.

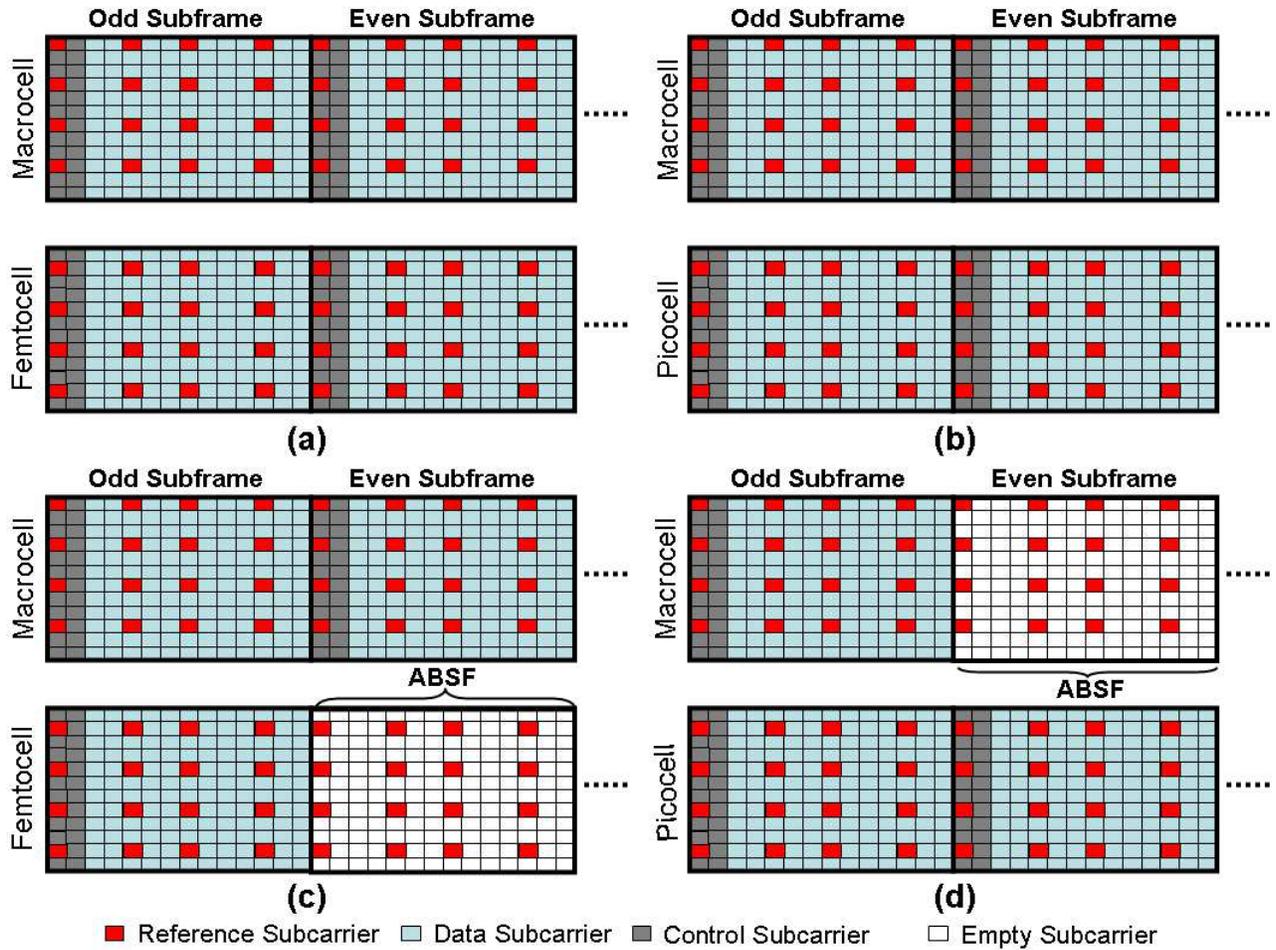

Fig. 2. Illustration of ABSFs for time-domain eICIC for 3GPP heterogeneous networks: (a) Macrocell and femtocell subframes without any eICIC. (b) Macrocell and picocell subframes without any eICIC. (c) Macrocell and femtocell subframes with eICIC. This can address the interference problem in Fig. 1(b), where the MUEs close to the femtocell can be scheduled in the even subframes of macrocell. (d) Macrocell and picocell subframes with eICIC. This can address the interference problem in Fig. 1(d), where the range-expanded PUEs can be scheduled in the even subframes of picocells.

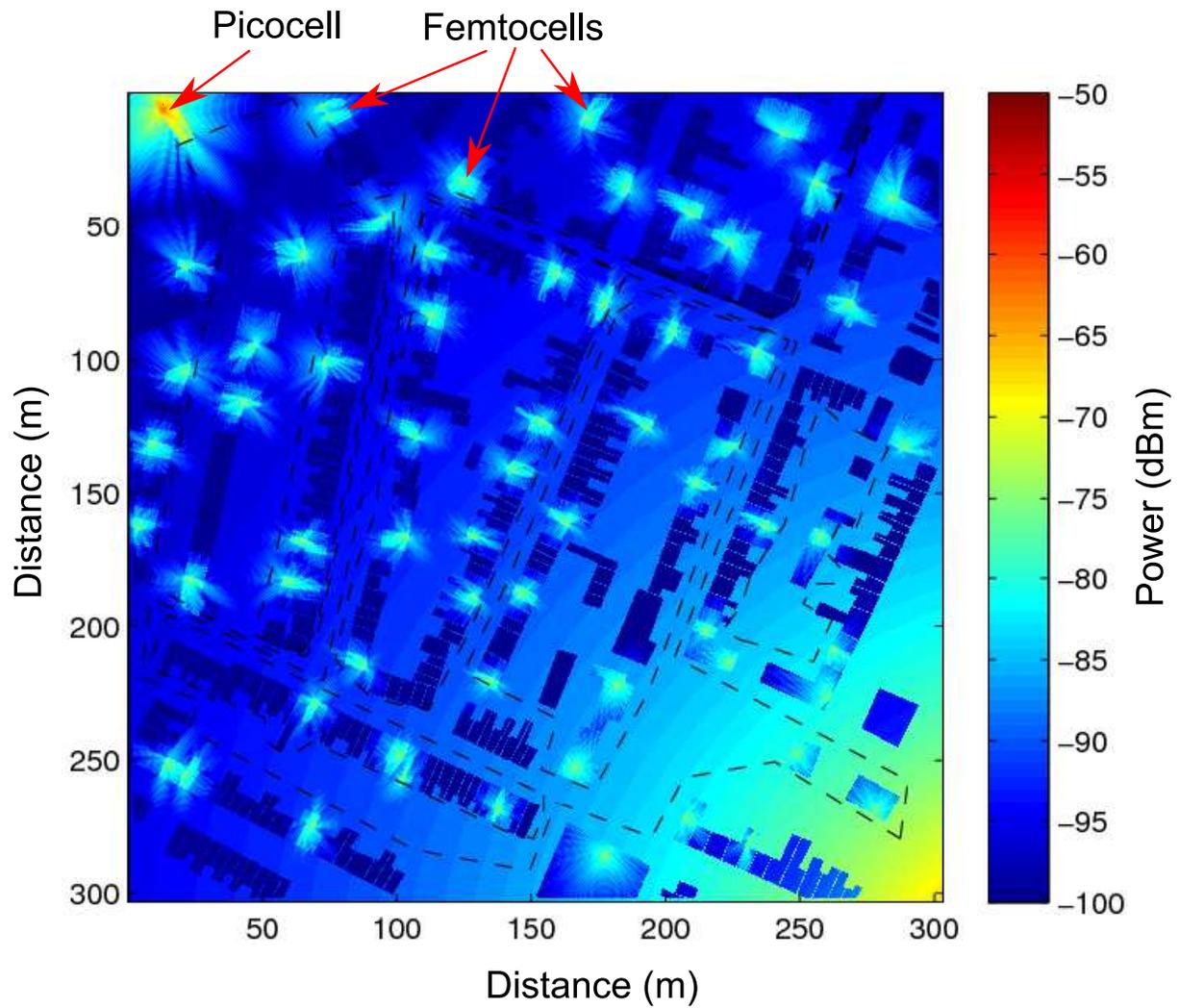

Fig. 3. HetNet simulated scenario based on an LTE-A network of 20 MHz with 1 macrocell transmitting at 46 dBm, 1 picocell transmitting at 30 dBm and 63 femtocells using up to 20 dBm. The dashed lines represent the routes followed by the eight macrocell users.

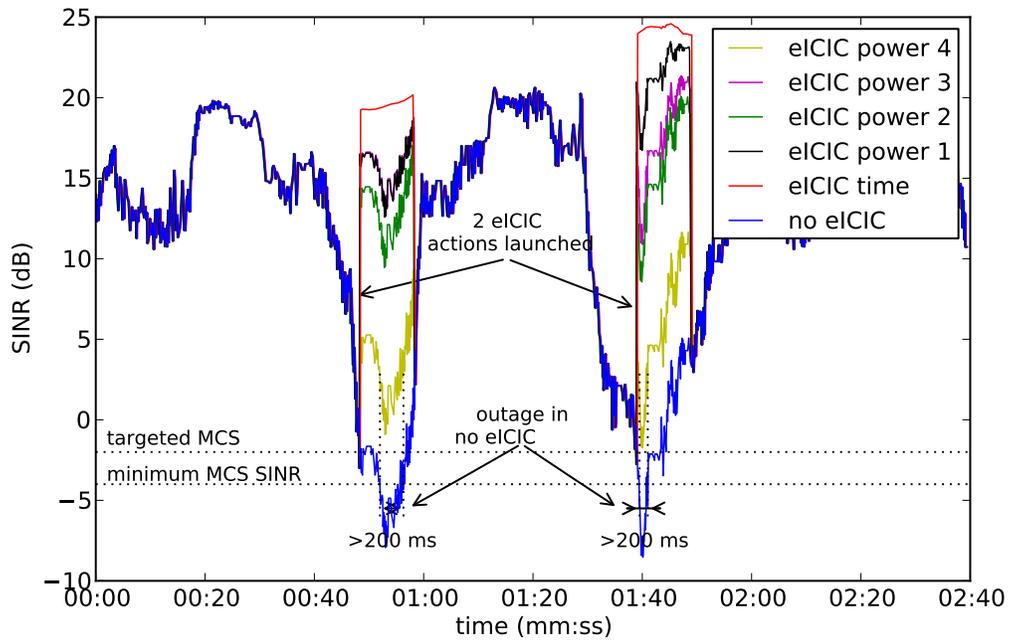

Fig. 4. SINR versus time of a victim MUE when passing close to two houses hosting a CSG femtocell.